# How Many Solar Neutrino Experiments Are Wrong?[*]

John N. Bahcall

*Institute for Advanced Study, School of Natural Sciences, Princeton, NJ 08540*

## Abstract

Ten recently-published solar models give $^7$Be neutrino fluxes that lie within a range of $\pm 10\%$ of the average value, a convergence that is independent of uncertainties in the measured laboratory rate of the $^7$Be$(p,\gamma)^8$B reaction. If nothing happens to solar neutrinos after they are created (*a la* standard electroweak theory) and the operating solar neutrino experiments are correct, then the $^7$Be solar neutrino flux must be less than 50% of the solar model value. At least three of the four existing solar neutrino experiments must be wrong *if*: (1) standard electroweak theory is correct, and (2) the true $^7$Be neutrino flux lies within the range predicted by standard solar models.

## I. INTRODUCTION

The purpose of this letter is to demonstrate that the solar neutrino problem can not be "solved" by postulating that only one or two experiments are incorrect. The focus of the argument presented here is the deficit of $^7$Be neutrinos that is revealed by a comparison of the chlorine and the Kamiokande solar neutrino experiments and independently by the results of the GALLEX and SAGE (gallium) solar neutrino experiments.

I use only the published results of the four ongoing solar neutrino experiments and the most robustly predicted neutrino fluxes from published standard solar models. Specifically,

---

the considerations developed here are independent of all uncertainties in the standard solar model calculations of the $^8$B neutrino flux. As is well known, the uncertainties in the theoretical calculations are largest for the $^8$B solar neutrino flux, in part because of difficulties in measuring precisely the laboratory rate for the $^7$Be$(p, \gamma)^8$B reaction.

A number of authors have presented arguments previously that the solar neutrino problem could not be solved by changing the solar model predictions [1–5]. These authors have largely concentrated on the comparison of the results from the chlorine and Kamiokande experiments, whereas I show here that the gallium discrepancy with the standard model is comparable in seriousness to the discrepancy between the chlorine and the Kamiokande results. The most important new ingredient used in the present investigation is the assumption that the rate of the fundamental $pp$ reaction, which is believed to be responsible for more than 98% of the solar luminosity, is given correctly by the standard solar model.

I first discuss in §II the inferences that follow from comparing the measured rates of the chlorine and the Kamiokande solar experiments. Then I compare in §III the results of the gallium experiments (GALLEX and SAGE) to the predictions of the standard solar model. Finally, in §IV I summarize the results and discuss how many solar neutrino experiments must be wrong if nothing happens to solar neutrinos after they are created and if the robust predictions of the standard solar model are correct.

I use throughout this paper the conservative procedures of the Particle Data Group [6] for estimating confidence limits.

## II. CLORINE VERSUS KAMIOKANDE

Let us begin by comparing the observed rates of the chlorine and of the Kamiokande solar neutrino experiments. The most recent result of the Kamiokande experiment is [7]

$$\phi_K(^8\text{B}) = \left(2.89^{+0.22}_{-0.21} \pm 0.35(\text{syst})\right) \times 10^6 \text{ cm}^{-2}\text{s}^{-1}, \tag{1}$$

where the first errors on the right hand side of Eq. (1) are statistical in origin and the additional errors are systematic. If standard electroweak theory is correct, then the shape of the energy spectrum from $^8$B solar neutrinos must be the same as the shape determined from laboratory experiments with terrestrial sources [8]. The cross section for $^8$B neutrinos with the standard energy spectrum incident on a $^{37}$Cl nucleus is [9] $\sigma_{Cl}(^8\text{B}) = 1.11(1.0 + \pm 0.025) \times 10^{-42}$ cm$^2$. Therefore, the capture rate in the chlorine experiment just from $^8$B neutrinos that are observed in the Kamiokande experiment is, if the energy spectrum is unchanged:

$$\phi_K(^8\text{B})\sigma_{\text{Cl}} = \left(3.21^{+0.24}_{-0.23} \pm 0.39(\text{syst})\right) \text{ SNU}, \tag{2}$$

where 1 SNU = $10^{-36}$ captures per target atom per sec.

The observed rate in the chlorine experiment [10] from all neutrino sources is

$$\text{Rate Cl Obs} = (2.55 \pm 0.17(\text{stat}) \pm 0.18(\text{syst})) \text{ SNU}. \tag{3}$$

Subtracting the rate due to $^8$B neutrinos that are observed in the Kamiokande experiment (Eq. 2) from the rate due to all neutrino sources (Eq. 3), one finds that the best estimate for the capture rate in the chlorine experiment from all sources except $^8$B neutrinos is

$$\text{Rate Cl Obs}\left(pep + {}^7\text{Be} + \text{CNO}\right) = (-0.66 \pm 0.52) \text{ SNU}, \tag{4}$$

if standard electroweak theory is correct. In deriving Eq.(4), I assumed that statistical and systematic errors are independent and that systematic errors in the two experiments are uncorrelated. I therefore combined the errors quadratically (see Particle Data Group [6]).

The negative value given in Eq. (4) for the sum of the capture rates from three different neutrino sources is the simplest expression of the "solar neutrino problem" and is independent of any theoretical parameters or characteristics of the solar model (cf. [1]).

Although the best estimate for the residual capture rate given by Eq. (4) is negative, the physical capture rate for any set of neutrino fluxes is positive definite. A conservative upper limit for the summed capture rates of *pep*, $^7$Be, and CNO solar neutrinos can be established by noting that 95% of the area in the physical region of the fluxes satisfies

$$(\phi\sigma)(pep) + (\phi\sigma)(^7\text{Be}) + (\phi\sigma)(\text{CNO}) \leq 0.68 \text{ SNU}, 95\% \text{ conf. limit Cl expt.}. \tag{5}$$

This Bayesian statistical procedure (with a flat prior) is the one adopted by the Particle Data Group [6].

We can refine Eq. (5) by utilizing the fact that the flux of neutrinos from the *pep* reaction ($p + e^- + p \rightarrow{}^2H + \nu_e$) is directly related to the basic *pp* reaction ($p + p \rightarrow{}^2H + e^+\nu_e$), which is the initiating fusion reaction for the fusion reaction that produces nearly all of the solar luminosity in standard solar models [11]. The estimated $1\sigma$ uncertainty for the *pep* neutrino flux in the standard model is only about 1.5 % [11,12]. This fact is illustrated in Table 1, which shows in the second column the results calculated for the *pep* capture rate from ten separate solar models with a wide variety of physics input and calculational procedures. The total spread in the calculated capture rates is

$$(\phi\sigma)_{\text{Cl}}(\text{pep}) = (0.22 \pm 0.01) \text{ SNU}, \quad \text{standard models}, \tag{6}$$

where much of the dispersion is due to round-off errors in the published *pep* fluxes. Subtracting the accurately-known standard model *pep* flux from the the upper limit for the three neutrino sources shown in Eq. (5), I obtain:

$$(\phi\sigma)(^7\text{Be}) + (\phi\sigma)(\text{CNO}) \leq 0.46 \text{ SNU}, \ 95\% \text{ conf. limit Cl expt.}. \tag{7}$$

The $^7$Be neutrino flux is predicted with moderate accuracy, $\pm 6$ % uncertainty [11,12]. The results from the ten models shown in Table 1 yield

$$(\phi\sigma)_{\text{Cl}}(^7\text{Be}) = (1.1 \pm 0.1) \text{ SNU}, \quad \text{standard models}. \tag{8}$$

The discrepancy between Eq. (7) and Eq. (8) is a quantitative statement of the solar neutrino problem that results from the chlorine and the Kamiokande experiments. The upper limit on the sum of the capture rates from $^7$Be and CNO neutrinos given in Eq. (7) is significantly less than the lowest value predicted for $^7$Be neutrinos alone, Eq. (8), by any of the ten recent standard solar models. I did not subtract the CNO neutrino capture rate from the sum of the two rates, Eq. (7), because the conflict between the measurements and the standard models (solar and electroweak) is apparent without this additional step and because the estimated rate [11,12] from CNO neutrinos, $0.4 \pm 0.08$ SNU, is more uncertain than for the other neutrino fluxes we are considering.

The flux of $^7$Be neutrinos is independent of uncertainties in the measurement of the low-energy cross section for the $^7$Be$(p,\gamma)^8$B reaction, the most uncertain quantity in the solar model calculations. One can establish this independence by recalling (see [11]) that the flux of $^7$Be neutrinos depends upon the proton-capture reaction only through the ratio

$$\phi(^7\text{Be}) \propto \frac{\text{R}(e)}{\text{R}(e) + \text{R}(p)}, \tag{9}$$

where $R(e)$ is the rate of electron capture by $^7$Be nuclei and $R(p)$ is the rate of proton capture by $^7$Be. With standard parameters, solar models give $R(p) \approx 0.001 R(e)$. Therefore, one would have to increase the value of the $^7$Be$(p,\gamma)^8$B cross section by more than two orders of magnitude over the current best-estimate [13] (which has an estimated uncertainty of less than 10%) in order to affect significantly the calculated $^7$Be solar neutrino flux.

### III. GALLEX AND SAGE

The GALLEX [14] and the SAGE [15] gallium solar neutrino experiments have reported consistent neutrino capture rates (respectively, $79 \pm 11.7$ SNU and $73 \pm 19.3$ SNU). Forming the weighted average of their results, the best estimate for the gallium rate is

$$\text{Rate Ga Obs} = (77 \pm 10) \text{ SNU}. \tag{10}$$

All standard solar models yield essentially the same predicted event rate from $pp$ and $pep$ neutrinos, i. e. (cf. Table 1),

$$(\phi\sigma)_{Ga}(\text{pp}) + (\phi\sigma)_{Ga}(\text{pep}) = (74 \pm 1) \text{ SNU}, \quad \text{standard models}. \tag{11}$$

Subtracting the rate from $pp$ and $pep$ neutrinos from the total observed rate, one finds that the combined rate from $^7$Be and $^8$B solar neutrinos in gallium is small,

$$\text{Rate Ga}(^7\text{Be} + ^8\text{B}) = (3 \pm 10) \text{ SNU}. \tag{12}$$

This result implies that

$$(\phi\sigma)_{\text{Ga}}(^7\text{Be}) + (\phi\sigma)_{\text{Ga}}(^8\text{B}) \leq 22 \text{ SNU}, \quad 95\% \text{ conf. limit}. \tag{13}$$

The combined $^7$Be and $^8$B rate given in Eq. (13) is less than the predictions from $^7$Be neutrinos alone for all 10 standard solar models shown in Table 1.

Moreover, one should take account of the $^8$B flux that is observed in the Kamiokande experiment [7], which in the gallium experiments amounts to

$$\phi_K \sigma_{\text{Ga}}(^8\text{B}) = 7.0^{+7}_{-3.5} \text{ SNU}, \tag{14}$$

where the quoted errors are dominated by the uncertainties in the excited state transitions in gallium [11] and I assumed that the shape of the energy spectrum of $^8$B solar neutrinos is the same as the laboratory shape [8]. Subtracting the rate from $^8$B neutrinos, Eq. (14), from the combined rate, Eq. (12), of $^7$Be and $^8$B neutrinos, one again finds a negative best-estimate flux for $^7$Be neutrinos,

$$(\phi\sigma)_{\text{Ga}}(^7\text{Be}) = (-4^{+11}_{-12}) \text{ SNU}. \tag{15}$$

The negative flux for the $^7$Be neutrinos that is inferred from the results of the GALLEX and SAGE experiments is the simplest version of the gallium solar neutrino problem.

Following the same statistical procedure as described earlier, one can set a conservative upper limit on the $^7$Be neutrino flux using the gallium and the Kamiokande measurements. One finds,

$$(\phi\sigma)_{\text{Ga}}(^7\text{Be}) \leq 19 \text{ SNU}, 95\% \text{ conf. limit}. \tag{16}$$

The predicted rate given by the 10 standard models listed in Table 1 is

$$(\phi\sigma)_{\text{Ga}}(^7\text{Be}) = (34 \pm 4) \text{ SNU}, \quad \text{standard models}. \tag{17}$$

The discrepancy between Eq. (16) and Eq. (17) is a quantitative expression of the gallium solar neutrino problem.

## IV. DISCUSSION AND CONCLUSION

The principal conclusions of this letter are listed below.

(1) The $^7$Be neutrino flux is robustly predicted by standard solar models (see [11] and the results summarized in Table 1 for 10 recently-published standard solar models). The calculated $^7$Be neutrino flux is independent of uncertainties associated with the difficult-to-measure laboratory cross section for the $^7$Be$(p,\gamma)^8$B reaction (see discussion following Eq. 9 ).

(2) The best-estimate for the combined neutrino events from $pep$, $^7$Be, and CNO neutrinos is negative, ( $-0.66 \pm 0.52$ SNU, see Eq. 4 ), if one uses the observed counting rates in the chlorine and the Kamiokande experiments and assumes that nothing happens to solar neutrinos after they are created. The inferred 95% confidence upper limit on the sum of the $^7$Be and CNO contributions to the chlorine event rate, 0.46 SNU, is less than half the predicted rate, $1.1 \pm 0.1$ SNU, from $^7$Be alone.

(3) Considering only the GALLEX and SAGE experiments, the 95% confidence upper limit on the sum of the $^7$Be rate and $^8$B rates, 22 SNU, is 65% of the rate predicted for $^7$Be alone, $34 \pm 4$ SNU(see Eq. 13. and Eq. 17 ). If one subtracts the $^8$B flux inferred from the Kamiokande experiment, the best estimate for the $^7$Be rate is again negative (see Eq. 15 ) and the 95% confidence upper limit is 19 SNU (Eq. 16).

There are therefore two solar neutrino problems: the chlorine-Kamiokande solar neutrino problem and the gallium solar neutrino problem.

Let us assume for purposes of discussion that a correct solar neutrino experiment must yield a rate for the $^7$Be neutrino flux that is consistent at the 95% confidence limit with nothing happening to solar neutrinos after they are created and with the value of the $^7$Be neutrino flux that is predicted by the standard solar model. If these assumptions are both correct, then at least three of the four operating solar neutrino experiments must be wrong. Either the chlorine or the Kamiokande experiment must be wrong in order to avoid conclusion (2) above and the GALLEX and SAGE experiments must be wrong in order to avoid conclusion (3) above. On the other hand, if one accepts the MSW description of neutrino propagation [24,25], all of the operating experiments can be fit well assuming also the correctness of the standard solar model.

What will the BOREXINO experiment see? BOREXINO [26] will provide the first direct measurement of the $^7$Be solar neutrino flux. According to the above-described arguments, the $^7$Be flux must be at least a factor of two lower than the standard solar model prediction if standard electroweak theory is correct and the rates measured by existing solar neutrino experiments are accurate. Even if the MSW theory is the correct explanation of the discrepancies between solar model predictions and experiments, the $^7$Be flux must be approximately a factor of two or more less than the solar model value [27].

In an insightful analysis, Kwong and Rosen [28] were the first authors to stress the centrality of the depletion of the $^7$Be neutrino flux to the solar neutrino problem. They generalized the discussion of Bahcall and Bethe [1] to the case in which oscillations among neutrinos are allowed. Kwong and Rosen emphasize that even when oscillations are assumed to occur the observed $^7$Be neutrino flux appears to be depleted. The arguments in the present paper differ from the Kwong-Rosen analysis in assuming that no oscillations occur, in adopting a more conservative statistical procedure, and in assuming the $p-p$ neutrino flux is predicted correctly by the standard solar model[1]. Using the conservative statistical procedure adopted here and allowing for neutrino oscillations, the Kwong and Rosen 95% upper limit on all but $^8$B neutrinos for chlorine becomes 1.13 SNU.

The ideas discussed in this paper were described in large part in invited talks in May 1994 at the Neutrino '94 conference in Eilat, Israel, the PASCHOS conference in Syracuse, New York, and the GONG '94 conference in Los Angeles. I am grateful to the organizers and participants in these conferences for valuable suggestions and comments, which sharpened the arguments. I am especially indebted to S. P. Rosen for an early copy of the important Kwong and Rosen paper, to H. Frisch and D. E. Groom for discussions of the relevant statistical procedures of the Particle Data Group, and to T. Bowles, A. Gould, H. Harari, K. Lande, T. Kirsten, Y. Totsuka, and S. Tremaine for stimulating discussions. This work was supported in part by NSF grant PHY-92-45317 with the Institute for Advanced Study.

---

[1]Kwong and Rosen estimate confidence limits by integrating the normal distribution over both negative and positive neutrino fluxes and by adding the estimated uncertainties to a negative best-estimate.

TABLES

TABLE I. Neutrino Rates from Recent Standard Solar Models. The entries are in SNU ($10^{-36}$ captures per target atom per second). The models are described in [4,12,16–23]. The two entries from [12] refer, respectively, to models computed with and without helium diffusion.

| Source | Cl | | Ga | |
| --- | --- | --- | --- | --- |
| | pep | $^7$Be | pp + pep | $^7$Be |
| Bahcall and Ulrich 1988 | 0.22 | 1.13 | 74 | 34 |
| Sackman et al. 1990 | 0.21 | 1.02 | 74 | 31 |
| Bahcall and Pinsonneault 1992 | 0.23 | 1.11 | 74 | 34 |
| Bahcall and Pinsonneault 1992 | 0.23 | 1.17 | 74 | 36 |
| Castellani et al. 1994 | 0.22 | 1.10 | 75 | 36 |
| Turck Chièze and Lopes 1993 | 0.22 | 1.02 | 74 | 31 |
| Berthomieu et al. 1993 | 0.22 | 1.07 | 73 | 33 |
| Castellani et al. 1994 | 0.22 | 1.15 | 74 | 35 |
| Kovetz and Shaviv 1994 | 0.22 | 1.18 | 74 | 36 |
| Proffitt 1994 | 0.22 | 1.24 | 73 | 38 |
| Christensen-Dalsgaard 1994 | 0.23 | 1.12 | 74 | 34 |